# Identifying and extracting Data Access Statements from full-text academic articles

*Matteo Cancellieri, CORE, KMi, The Open University. matteo.cancellieri@open.ac.uk*

*David Pride, CORE, KMi, The Open University. david.pride@open.ac.uk*

*Petr Knoth, CORE, KMi, The Open University. petr.knoth@open.ac.uk*

## Abstract

A Data Access Statement (DAS) is a formal declaration detailing how and where the underlying research data associated with a publication can be accessed. It promotes transparency, reproducibility, and compliance with funder and publisher data-sharing requirements. Funders such as Plan S, the European Union, UKRI, and NIH emphasise the inclusion of DAS in publications, underscoring its growing importance.

While a DAS enhances research by increasing transparency, discoverability, and data quality while clarifying access protocols and elevating datasets as first-class research outputs, the repository community faces challenges in managing and curating DAS as a standard metadata component. Manual DAS curation remains labour-intensive and time-consuming, hindering efficient data-sharing practices.

CORE has co-designed with the repository community a module that uses machine learning to identify and extract DAS from full-text articles. This tool facilitates the automated encoding, curation, and validation of DAS within metadata, reducing manual workload and improving metadata quality. This integration aligns with CORE's objective to enhance repository services by providing enriched metadata and supporting compliance with funder requirements. By streamlining DAS management and expanding metadata frameworks, CORE contributes to a more accessible and interconnected scholarly ecosystem, fostering data discoverability and reuse.

## Acknowledgements

We acknowledge the support of all the members of the CORE team in the development of the DAS integration in CORE and the CORE Dashboard.

## Introduction
A Data Access Statement (DAS) constitutes a formal declaration by an author describing how and where the underlying research data can be accessed. It ensures transparency, reproducibility, and compliance with funder or publisher requirements regarding data sharing.

Many funders bodies such as Plan S[1], the European Union[2], UKRI[3] and NIH[4] are all detailing the requirements of including Data Access Statements in publications with various degrees of strictness.

Data access statements offer numerous advantages for researchers, publishers, funders, and the wider scientific community.

- Increased transparency by providing detailed descriptions of how data can be accessed, enabling other researchers to verify findings, replicate studies, and build upon existing work. This facilitates reproducibility and strengthens the credibility of research.

- Increased discoverability, by including persistent identifiers, such as DOIs and connecting research papers to datasets, both the paper and the data is made more accessible within repositories and indexing platforms.

- Data quality improvements, by ensuring that datasets are published and released with the aim of maximum utility and reuse.

- Clarity on data access, helping users understand how to obtain and use the data responsibly.

- Raising the dataset status as first-class research outputs, the proper citation and acknowledgement facilitate citations, monitoring and collaboration across different disciplines and institutions.

Despite the numerous benefits, managing and capturing these statements as a standard component of the metadata associated is still a considerable challenge from the repository community perspective.

Many repository deposit pipelines allow for a manual addition of a Data Access Statement, however, the monitoring and curation of the statement is still a labour-intensive and time-consuming task, which may create obstacles for repository managers and hinder efficient data sharing management.

## Automatic extraction of data access statements

Following the direction and requests of CORE Members[3], CORE started co-designing a new CORE Dashboard module to support DAS discovery and management. The CORE team worked closely with repositories and research managers to understand their requirements and incorporate their feedback into the final product.

Based on the experience from the Rights Retention Statement module, CORE developed a new tool that employs a machine learning (ML) model to automatically identify and extract Data Access Statements from full-text articles, thereby streamlining the encoding, curation and validation of this information within the metadata.

Data Access Statements have been widely adopted over the past few years, with numerous institutional recommendations available on how to include them. As a result, DAS often follow common patterns and use standardized language. CORE's model leverages a relatively simple keyword-matching approach on the full text of papers to identify those potentially containing a DAS. The keywords were selected by analyzing a broad sample of papers across various fields, with the most frequently used terms assigned higher scores to improve the likelihood of selecting relevant papers. This straightforward and cost-effective method narrows the search set for locating DAS. Once this subset is identified, a more sophisticated scoring system, based on a combination of regular expressions, determines which papers genuinely include a DAS and extracts the corresponding dataset URL.

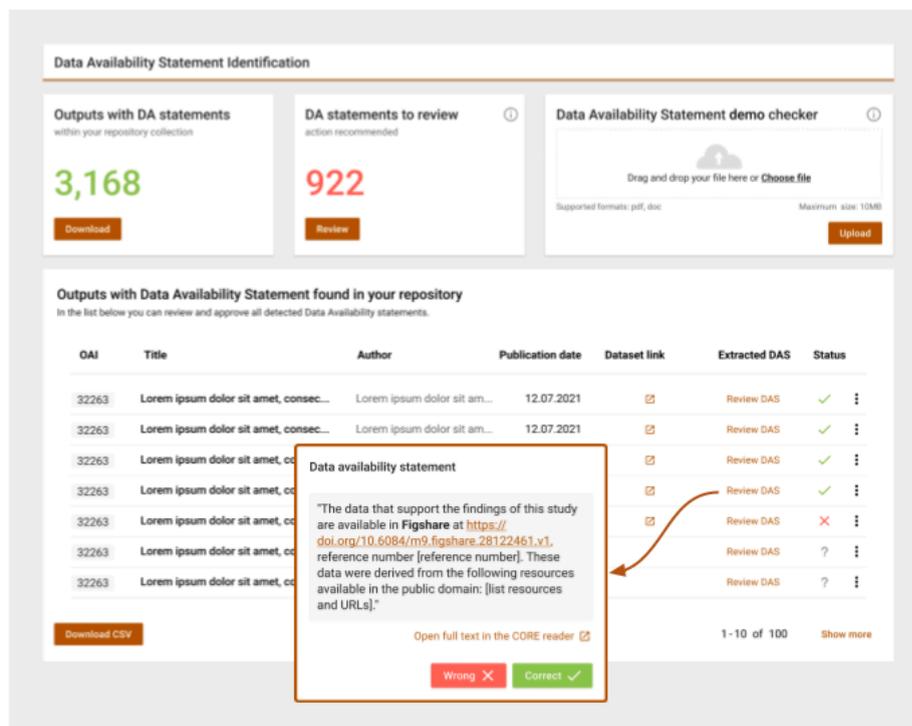

Figure 1: The Data Access Statement Tab wireframe

Once extracted, the DAS will be displayed in the Data Access Statement tab within the dashboard, allowing repositories to browse and manage the DAS directly or download a CSV for automated processing. Additionally, an option is available for uploading and automatically checking PDFs not included in the CORE collection. This feature supports repository managers in pre-deposit workflows and other activities involving papers not yet available in the collection.

In the current ecosystem, CORE has access to full text only after the record has been deposited and the metadata has been already recorded therefore such a module can be only used as a correction and cleaning tool rather than a proactive tool. Working together with repository software developers, CORE also is developing an additional way of accessing the new Data Access Statements identification tool through its API, allowing repository software to integrate the DAS matching at the time of deposit. This approach would enable the automatic analysis of the author's manuscript PDF and the matching and extraction of the DAS statement and also other enrichments such as the Rights Retention Statement if available. The extracted information can then be validated by the repository and will become part of the metadata record, reducing the manual burden on repository staff and improving consistency in capturing this information.

The closer integration of CORE services with repository software has been a longstanding objective, as it would significantly enhance the value of the data that CORE can provide to repositories.

Expanding the framework to incorporate additional data from manuscripts allows repositories to enhance metadata quality and increase its overall value. This comprehensive approach to data extraction and metadata management equips repositories to better meet the diverse needs of funders, institutions, and other stakeholders, fostering a more accessible and interconnected scholarly ecosystem. Integrating CORE services with repository software and broadening data extraction capabilities are crucial steps toward realizing this vision and fully leveraging the potential of research metadata.

# References (if applicable)

[1] Plan S Technical Guidance and Requirements
https://www.coalition-s.org/technical-guidance_and_requirements/

[2] Horizon Europe Programme Guide

https://ec.europa.eu/info/funding-tenders/opportunities/docs/2021-2027/horizon/guidance/programme-guide_horizon_en.pdf

[3]  UKRI: Making your research article Open Access
https://www.ukri.org/manage-your-award/publishing-your-research-findings/making-your-research-article-open-access/

[4] NIH Policy for Data Management and Sharing
https://grants.nih.gov/grants/guide/notice-files/NOT-OD-21-013.html

[5]Asking CORE members what matters to them.
https://blog.core.ac.uk/2023/07/06/asking-core-members-what-matters-to-them/

Asking CORE members what matters to them…